\documentclass[titlepage, aps,prb,twocolumn]{revtex4}

\usepackage{graphicx}
\usepackage{amsmath}
\usepackage[normalem]{ulem}

\usepackage{physics}

\usepackage[pdftex,colorlinks=true]{hyperref}
\hypersetup{
linkcolor=blue,
citecolor = blue,
urlcolor=blue
}

\begin{document}

 
\setlength{\pdfpagewidth}{8.5in}
\setlength{\pdfpageheight}{11in}

\title{Random singlet-like phase of disordered Hubbard chains}

\author{Josephine Yu$^{1}$, Hong-Chen Jiang$^{2}$, Rahul Nandkishore$^{3,4}$, 
 S. Raghu$^{4,2}$}

\affiliation{$^{1}$ Department of Applied Physics, Stanford University, Stanford, CA 94305, USA \looseness=-1}
\affiliation{$^{2}$ Stanford Institute for Materials and Energy Sciences, SLAC National Accelerator Laboratory, Menlo Park, CA 94025, USA \looseness=-1}
\affiliation{$^{3}$ Department of Physics and Center for Theory of Quantum Matter, University of Colorado, Boulder, CO, 80309, USA \looseness=-1}
\affiliation{$^{4}$ Department of  Physics, Stanford University, Stanford, CA 94305, USA \looseness=-1}

\date{\today}

\begin{abstract}
Local moment formation is ubiquitous in disordered semiconductors such as Si:P, where it is observed both in the metallic and the insulating regimes. Here, we focus on local moment behavior in disordered insulators, which arises from short-ranged, repulsive electron-electron interactions. Using density matrix renormalization group 
and strong-disorder renormalization group 
methods, we study paradigmatic models of interacting insulators: one dimensional  Hubbard chains with quenched randomness. In chains with either random fermion hoppings or random chemical potentials, {\it  both at and away from half-filling},  we find exponential decay of charge and fermion 2-point correlations but power-law decay of spin correlations that are indicative of the random singlet phase. The numerical results can be understood qualitatively by appealing to the large-interaction limit of the Hubbard chain, in which a remarkably simple picture emerges.  
\end{abstract}
\maketitle


\section{Introduction} 
A fundamental challenge in condensed matter physics has been to understand  the implications of local moment formation in disordered electronic systems. Local moments have long been observed in disordered semiconductors, such as Si:P and Si:P,B, where they crucially affect the thermodynamic\cite{Paalanen1986,Paalanen1988} and dynamical\cite{Mahmood2021} properties in both the metallic and the insulating regimes \cite{Bhatt1988,Belitz1994,Miranda2005}. If the properties of the phases themselves are altered by local moments, it  follows that at least in principle, local moments can influence the universal behavior near metal-insulator quantum phase transitions\cite{Rosenbaum1980,Sarachik1998,Kravchenko2003}.  

Much work has been done in understanding magnetic excitations in the metallic regime of disordered systems\cite{Finkelstein1990,Belitz1994}.  In two spatial dimensions, for instance, 
there is a tendency towards a 
magnetic instability even in the weak disorder limit, far from a putative metal-insulator transition\cite{Finkelstein1983,Andreev1998,Chamon1999,Kamenev1999,Chamon2000,Nosov2020}. By contrast, since the early seminal work of  Bhatt, Lee and Co-workers\cite{Bhatt1982,Zhou2009}, considerably less attention has been devoted to local moment behavior in the insulating regime.   

Assuming the existence of local moments in the insulator, the theory of Bhatt and Lee \cite{Bhatt1982}  establishes the tendency towards random singlet formation, due to an exponentially broad distribution of  antiferromagnetic exchange interactions among the local moments. Later work by Bhatt and Fisher  \cite{Bhatt1992} pushed this picture further, into the metallic regime, arguing that local moments essentially decouple from the metallic electrons, due to vanishingly small Kondo temperatures. Nevertheless, it remains unclear how such behavior emerges from electrons in a random landscape, in the presence of short-ranged interactions.  

In this paper, we report some progress in this direction and analyze models of electrons in the presence of both short-range interactions and strong disorder.  Given our focus on the insulating state, we study one dimensional models, in which the tendencies towards insulating ground states are strongest. We are especially interested in the behavior away from half-filling, where at least microscopically, a description in terms of local moments alone is not justified {\it a priori}.  

Using a combination of density matrix renormalization group 
simulations and real space renormalization group techniques, we demonstrate that the Hubbard chain exhibits random singlet behavior both at and away from half-filling.  Our conclusion holds for both random potentials (site disorder) and random hoppings (bond disorder).  To make intuitive sense of our results, we appeal to the strong-interaction limit of the Hubbard model and account for our results in terms of spin-charge separation\cite{Ogata1990}: nearly-free holes exhibit Anderson localization, while spins experience random Heisenberg exchange, resulting in random singlet formation along the lines of Bhatt and Lee.  

\section{Model}
The simplest effective Hamiltonian governing electrons with  disorder and short-range interactions is the Hubbard model with randomness:
\begin{equation}
H =  -\sum_{i,\sigma} t_{i} (c_{i,\sigma}^\dagger c_{i+1,\sigma}  +\text{h.c.})+ \sum_i \mu_i n_i +  U\sum_i n_{i\uparrow}n_{i\downarrow},  \label{eq:hubbard_hamiltonian} 
\end{equation}
where $c^{\dagger}_{i \sigma}(c_{i \sigma})$ creates(destroys) an electron with spin $\sigma = \uparrow, \downarrow$, on lattice site $i$, the density operator on site $i$ is $n_i = \sum_{\sigma} c^{\dagger}_{i \sigma} c_{i \sigma}$.  
Onsite interactions are taken to be repulsive: $U > 0$, $t_i$ are   nearest-neighbor hopping amplitudes, and $\mu_i$ are on-site chemical potentials.  

\begin{figure*}
\begin{center}
\includegraphics[scale=1]{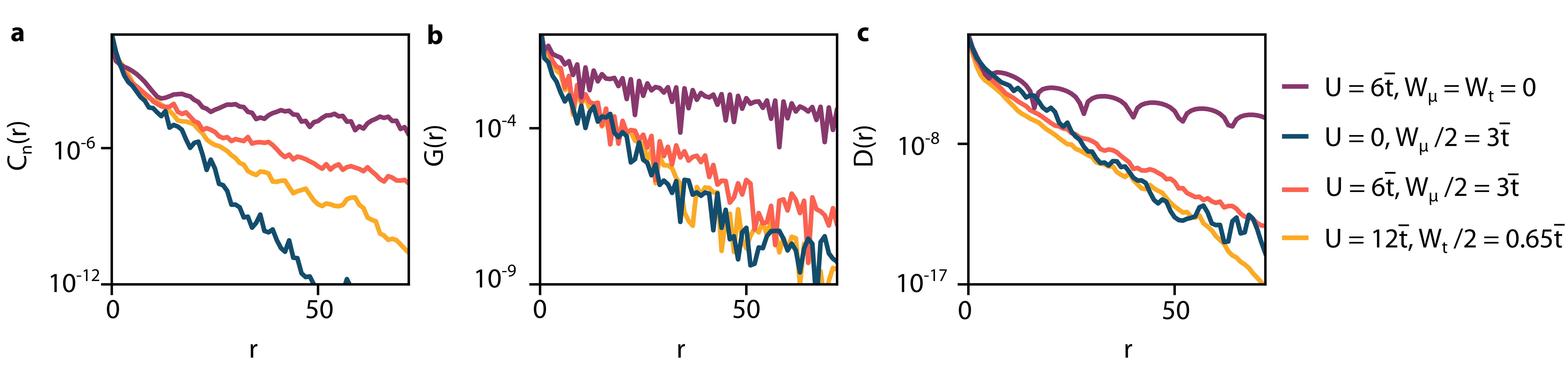}
\end{center}
\caption{ Density-density, fermion 2-point and pair-field correlation functions on $L=144$ Hubbard chains at electron filling $n=11/12$ with random potential, random hopping as well as for the disorder-free 
systems. Additionally, results for the non-interacting ($U=0$) but site-disordered system are shown. All data is shown for $r\leq L/2$. The random-potential system has parameters $U=6 \overline{t}$, $W_\mu/2 = 3\overline{t}$, $W_t = 0$; the random-hopping system has parameters $U=12\overline{t}$, $W_\mu =0$, $W_t/2 = 0.65\overline{t}$; the clean system has parameters $U=6\overline{t}$, $W_\mu = W_t = 0$; and the Anderson system has $U=0$, $W_\mu/2=3\overline{t}$, $W_t = 0$. (a) Disorder-averaged density-density correlations $\overline{C_n(r)}$ (Eq. \ref{eq:densitydensity}) decay exponentially with distance $r$ for the disordered systems, as opposed to decaying with a power law as in the clean system. (b)   Electron 2-point functions decay exponentially in all disordered systems. (c) Superconducting pair-field correlations decay exponentially in the disordered systems, in direct contrast to the power-law decay in the non-disordered Luttinger liquid.  \label{fig:charge}}
\end{figure*}

We study two types of quenched randomness: 
site disorder, where the local chemical potentials $\mu_i$ are random, and 
bond disorder, where 
$t_{i}$ are random. We choose $\mu_i$ from a uniform 
distribution of mean $\overline{\mu}=0$ and width $W_\mu$. Similarly, $t_i$ are chosen from a uniform 
distribution of mean $\overline{t}$ and width $W_t$. We choose $\overline{t}=1$, setting this to be the unit of energy. 


In the absence of randomness, the model is integrable and has of course been thoroughly studied\cite{Arovas2021}; the half-filled system is well-described by a $S=1/2$ Heisenberg antiferromagnet, and the doped chain exhibits Luttinger liquid behavior over a range of electron concentrations and interaction strengths. With perturbatively weak disorder, it is known that the Luttinger liquid tends towards localization and that repulsive interactions enhance this tendency \cite{Giamarchi1987}.  At half-filling, bond disorder preserves particle-hole symmetry, and a spin chain with random antiferromagnetic exchange accurately captures the low energy behavior of the system.  It has been well-established that the latter results in an infinite randomness fixed point with random singlet behavior \cite{Ma1979,Fisher1994}.  We wish to explore the fate of the Hubbard chain with site and bond randomness, both at and away from half-filling where such a description in terms of spin alone is not necessarily valid.

\section{DMRG Results}
We analyze Hubbard chains described by Eq. \ref{eq:hubbard_hamiltonian} at two fixed electron filling fractions, $n=1$ (half-filling) and $n=11/12$, using the density matrix renormalization group (DMRG)\cite{White1992,White1993} procedure. We perform all of the simulations in the strong-interaction regime, $U = 12\overline{t}$ for the random-hopping chains or $U=6\overline{t}$ for the random-potential chains. 
The DMRG algorithm works well for both clean and weakly disordered systems. For a more reliable study, we first obtain the ground state of a system with weak disorder, then quasi-adiabatically increase the disorder strength, adaptively increasing sweep number and the number of basis states kept, until the resulting well-converged ground state has been obtained. A similar procedure has been used before in DMRG to treat disordered systems \cite{Xavier2018}. In the present study, we perform up to 50 sweeps and keep up to $m=1024$ states with a typical truncation error $\epsilon \sim 10^{-9}$. For all parameters, we sample at least 300 independent disorder realizations.

To characterize the ground state properties, 
we calculate various equal-time correlation functions over the interior half of the chain (from site L/4 to 3L/4), 
to minimize the boundary effects. We focus on measures of the charge and spin behaviors, probed through the charge density-density fluctuation correlation function
\begin{equation}
C_n(r) \equiv \expval{(n(x)-\expval{n(x)})(n(x+r)-\expval{n(x+r)})}
\label{eq:densitydensity}
\end{equation}
and the spin-spin correlation function
\begin{equation}
C_\sigma(r) \equiv \expval{S(x)S(x+r)} \label{eq:spinspin}
\end{equation}
respectively, where $r$ is the displacement between two sites along the chain and $x = L/4$ is a fixed reference point.
We also measure the fermion 2-point function 
\begin{equation}
G(r) \equiv \expval{c_\uparrow^\dagger(x)c_\uparrow(x+r)}, \label{eq:singleptcle}
\end{equation}
where the choice of the up-spin does not matter due to the spin $SU(2)$ symmetry.

Lastly, the superconducting pair-field correlation function is defined as
\begin{equation}
D(r) \equiv \expval{\Delta^\dagger(x)\Delta(x+r)} \label{eq:sccor}
\end{equation}
where $\Delta(y)  \equiv (c_\downarrow^\dagger(y) c_\uparrow^\dagger(y)-c_\uparrow^\dagger(y) c_\downarrow^\dagger(y))/\sqrt{2}$ is the spin-singlet pair creation operator on bond $y$.

We start by presenting evidence for charge localization in the disordered systems away from half-filling. 
In the absence of disorder, the $n=11/12$ system has a Luttinger liquid ground state. We observe that disorder, regardless of type, localizes the charges. 
Fig. \ref{fig:charge} shows exponential decay of the disorder-averaged density-density correlation function $\overline{C_n(r)}$ (Eq. \ref{eq:densitydensity}), the fermion 2-point function $\overline{G(r)}$ (Eq. \ref{eq:singleptcle}), and the superconducting pair-field correlation function $\overline{D(r)}$ (Eq. \ref{eq:sccor}) of the ground state  in the disordered systems as a function of distance $r$, indicating a gap to charge excitations. Qualitatively, the behaviors of these correlation functions in the random-potential and random-hopping systems are very similar, both differing significantly from the power-law correlations expected for the disorder-free system. These correlation functions for both site and bond randomness resemble those of an Anderson insulator. 
\begin{figure}[h]
\begin{center}
\includegraphics[scale=1]{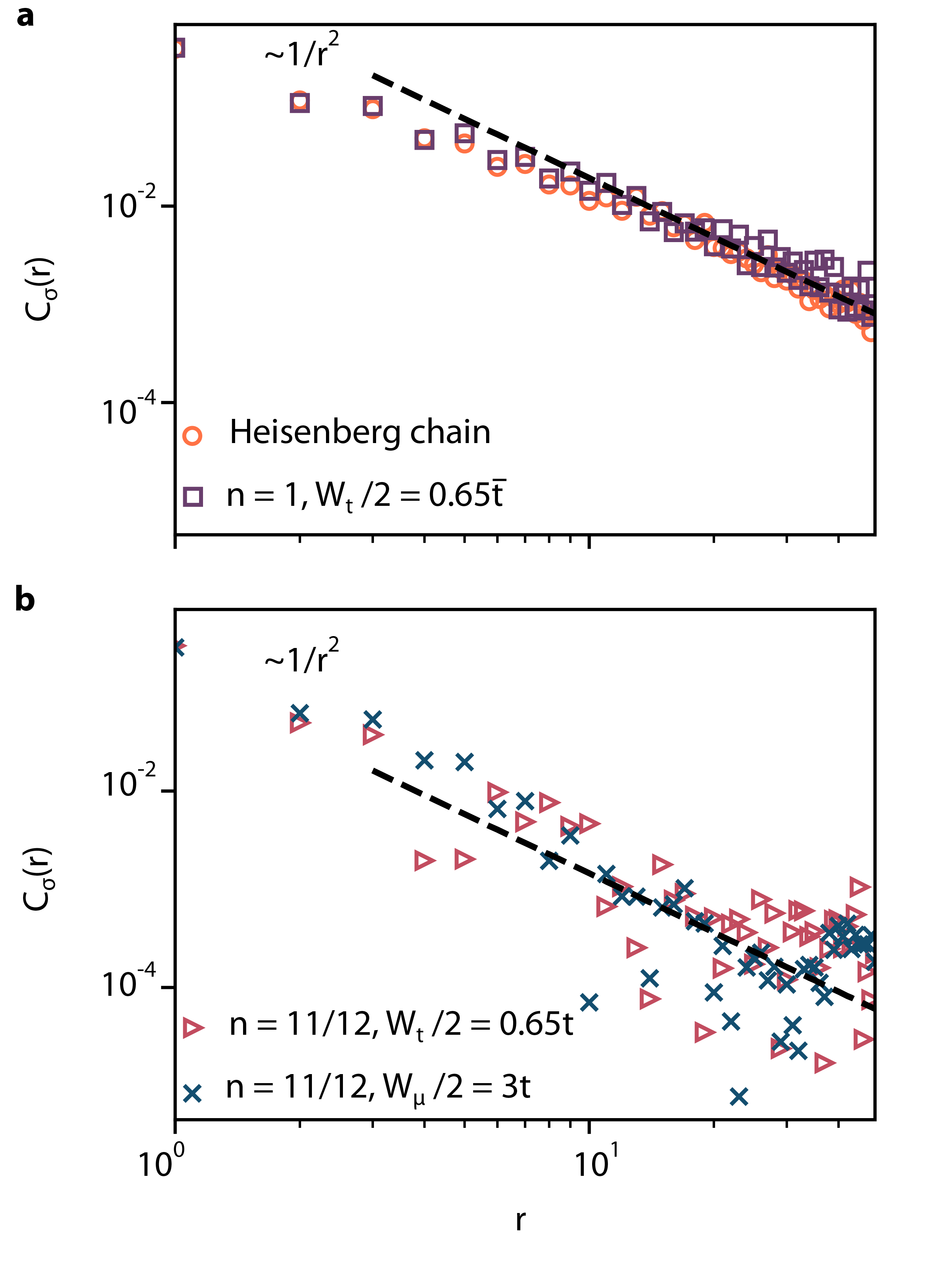} 
\end{center}
\caption{ Disorder-averaged spin-spin correlation $\overline{C_\sigma(r)}$ (Eq. \ref{eq:spinspin}) for $L=144$-site chains.  Data is shown for distances $r \leq L/3$ , as fluctuations increase with distance. Black dashed lines indicate $~r^{-2}$ decay. (a) Random Heisenberg chain with spin exchange couplings derived perturbatively from the Hubbard parameters (yellow) and random-hopping Hubbard chain at half-filling (purple) (see main text). In both chains, $\overline{C_\sigma(r)}$ exhibits long-distance $r^{-2}$ behavior beyond around 10 lattice sites. (b) Spin correlations in random-hopping and random-potential systems at $n=11/12$ electron filling also exhibit decay close to $r^{-2}$. Insets show $r^2\cdot \overline{C_\sigma(r)}$ exhibiting stable long-distance random-singlet-like behavior, and solid lines in the insets are average values of $r^2\cdot \overline{C_\sigma(r)}$ at long distances. \label{fig:spin_corr_raw}}
\end{figure}

We now turn to analyzing the disorder-averaged spin-spin correlation functions $\overline{C_\sigma(r)}$ (Eq. \ref{eq:spinspin}) for different chains: random-potential and random-hopping Hubbard chains as well as the random-exchange Heisenberg antiferromagnetic chain. Despite the presence of disorder, we find that the spin $SU(2)$ symmetry is preserved in $C_\sigma(r)$ of each disorder realization and therefore in the disorder-averaged correlation function $\overline{C_\sigma(r)}$.

The half-filled system is unsurprisingly a Mott insulator, since the repulsive interaction $U$ is the dominant energy scale. The effective low-energy description of the half-filled, large-interaction systems should then be equivalent to the random Heisenberg antiferromagnet. The results for the random Heisenberg chain and the random-hopping Hubbard chain at half-filling are shown in Fig. \ref{fig:spin_corr_raw}a, and the agreement between them reflects this intuition. Our results agree also with previous studies of the disordered Hubbard chain at half-filling \cite{Melin2006}. We note that the random-potential chain at half-filling exhibits some curious charge behavior at weak disorder that can be understood through a particle-hole transformation (see Appendix). At large distances $r$, the spin correlations in both the half-filled Hubbard chain with random hoppings and the random Heisenberg chain exhibit decays close to $1/r^2$, the expected behavior in a random singlet phase. More surprisingly, Fig. \ref{fig:spin_corr_raw}b shows that $\overline{C_\sigma(r)}$ in the random hopping and random potential systems {\it away} from half-filling decay at large distances $r$ as a power law close to $1/r^2$, indicating that the spin order, both at and away from half-filling, are random-singlet-like. Away from half-filling, statistical fluctuations decrease more slowly with sample number, as electron configurations must now be taken into account.

Another hallmark of the random singlet phase is that the physics is dominated by rare, long-range singlets. We further characterize the systems studied here by comparing the disorder-averaged spin-spin correlations $\overline{C_\sigma(r)}$ to the root-mean-square (RMS) spin-spin correlations $\sqrt{\overline{(C_\sigma(r))^2}}$. In the random-singlet phase, both $\overline{C_\sigma(r)}$ and $\overline{(C_\sigma(r))^2}$ are dominated by the probability of forming a singlet of length $r$, which scales as $1/r^2$ at large $r$. One then expects the disorder-averaged spin-spin correlations to scale as $1/r^2$ and the RMS spin-spin correlations to scale as $1/r$ in the random-singlet phase. 
By contrast, the two quantities should agree in the weak-disorder limit, in which rare-region effects can be ignored. 
Fig. \ref{fig:avg_vs_sq} shows $\overline{C_\sigma(r)}$ and $\sqrt{\overline{(C_\sigma(r))^2}}$ with behaviors consistent with random-singlet physics for a random-potential system away from half-filling. Similar rare region-dominated behavior is found for the random-hopping system.

\section{Large-Interaction Limit and Numerical SDRG}

Our results thus far can be understood qualitatively through the simple picture of the Bethe Ansatz solution of the (clean) Hubbard chain in the $U/t\rightarrow \infty$ limit. In this limit, the spins and charges are decoupled: the holes are free to order $t/U$, and the spins form a Heisenberg antiferromagnet on electron coordinates\cite{Ogata1990}. One could thus expect the decoupled spins and charges to respond independently to the disorder. Then, the holes undergo Anderson localization, while the spins form a random singlet on electron coordinates. 

Quantitatively, we can explore this perspective by numerically implementing a strong-disorder renormalization group (SDRG) decimation procedure. We again study the $n=11/12$ chain with random potential, imposing now that the holes are localized (as justified by the evidence from DMRG, Fig. \ref{fig:charge}) at local maxima in the potential of a given disorder realization so that we are left with an effective spin model. The effective spin model, a random Heisenberg chain on electron coordinates, can be computed using perturbation theory in $t/U$. We then numerically implement the SDRG decimation of the effective random spin model (see Appendix). Averaging the resulting ground states of many realizations is equivalent to averaging over all possible positions of the localized holes, and one recovers random singlet behavior.  

\begin{figure}[h]
\begin{center}
\includegraphics[scale=1]{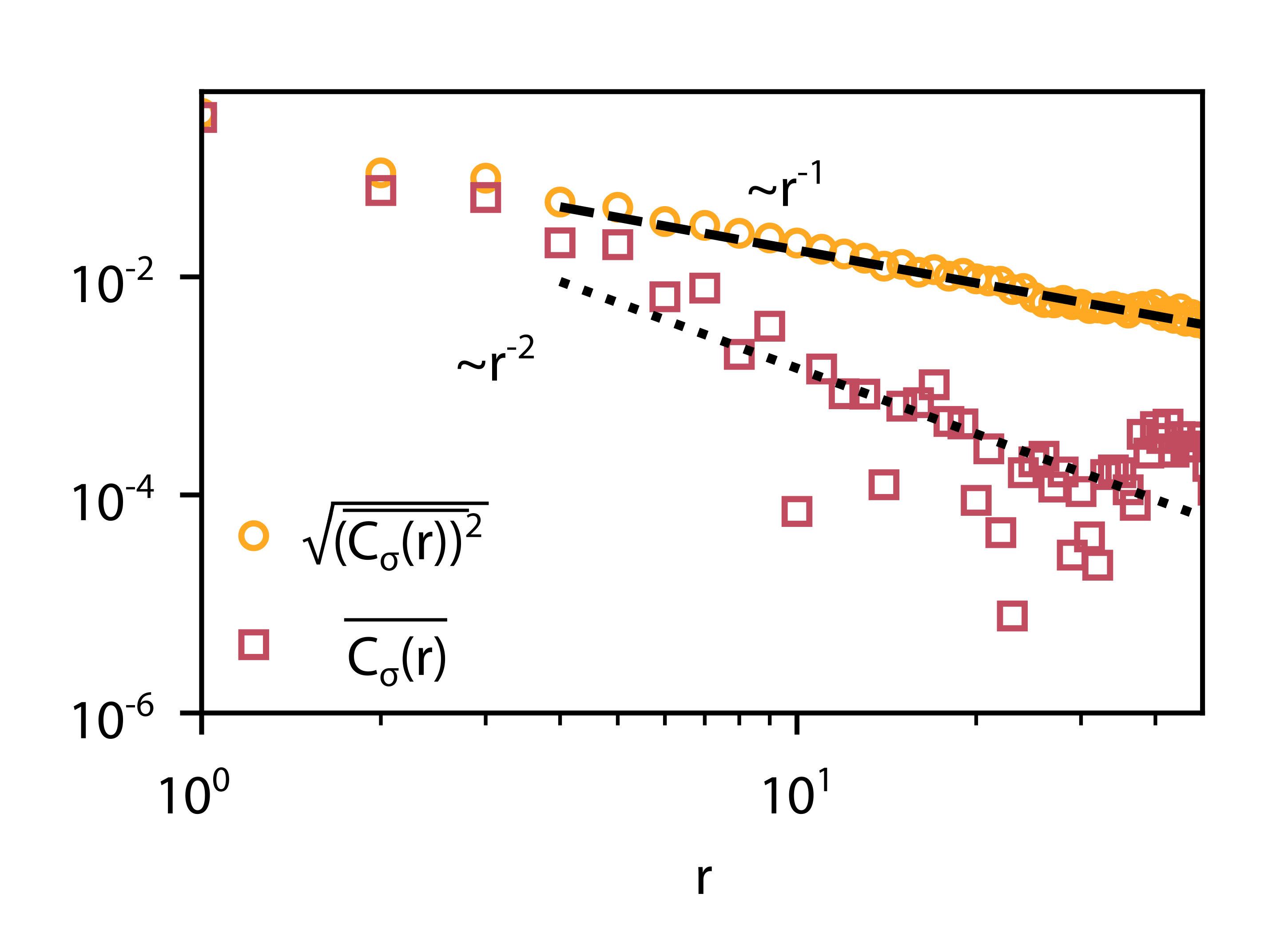} 
\end{center}
\caption{Spin correlation statistics in a $L=144$ site chain with random potential, $W_\mu/2 = 3\bar{t}$. Disorder-averaged spin correlations $\overline{C_\sigma(r)}$ and the root-mean-square spin correlations $\sqrt{\overline{(C_\sigma(r))^2}}$ shown for $r\leq L/3$. In the absence of disorder, the two quantities should be equivalent, but the   difference in  the two quantities indicates  the importance of rare regions indicative of a strong disorder fixed point. A dotted line shows a $1/r^2$ decay, and a dashed line shows a $1/r$ decay. \label{fig:avg_vs_sq}} 
\end{figure}

\begin{figure}[h]
\begin{center}
\includegraphics[scale=1]{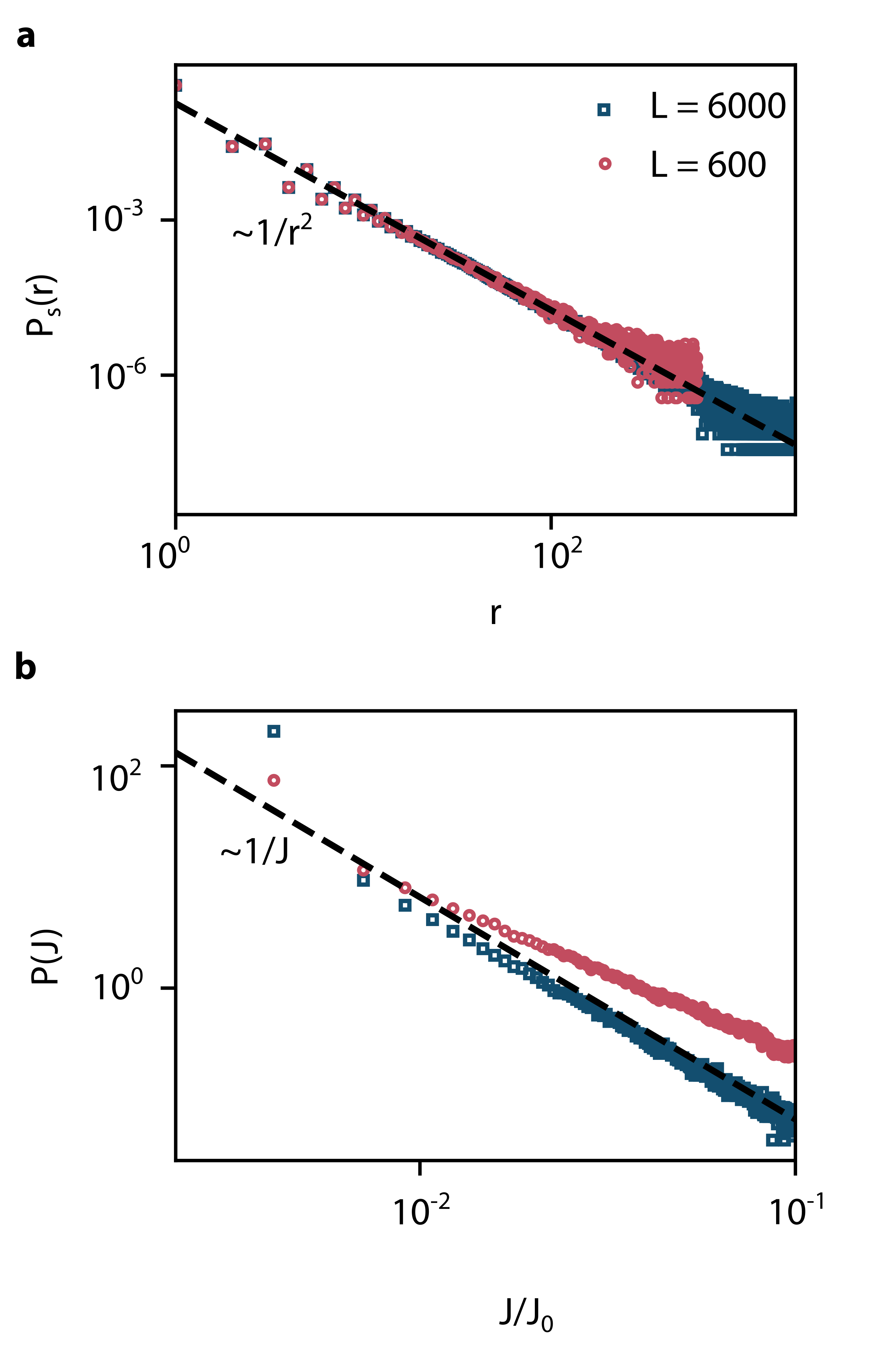} 
\end{center}
\caption{ Results of numerically implementing bond decimation, assuming charge localization, for a $L=600$-site chain and $L=6000$-site chain, both with electron density $n=11/12$, averaged over 5000 realizations. (a) Probability $P_s(r)$ of forming a singlet pair over $r$ lattice sites exhibits $r^{-2}$ scaling. Data is shown for $r\leq 2000$. (b) Probability distribution of $J$, spin exchange energies, in units of the energy scale $J_0 = \text{max}(\{J\})$ when $50$ free spins remain. The black line is a $1/J$ distribution normalized over the window of energies shown. There is strong agreement between the bond distribution found in the numerical SDRG procedure for the longer chain and the $1/J$ distribution of spin exchange energies predicted by the infinite-randomness fixed point corresponding to the random singlet phase.   \label{fig:numerical_rg}}
\end{figure}

From the numerical decimation, we obtain the probability $P_s(r)$ of forming a singlet pair separated by a distance of $r$ lattice sites, shown in Fig. \ref{fig:numerical_rg}a for a $L=600$-site chain and a $L=6000$-site chain at $n=11/12$ electron density. This probability distribution ultimately determines the long-distance behavior of the disorder-averaged spin-spin correlations $\overline{C_\sigma(r)}$\cite{Hoyos2007, Fisher1994}. The probability of forming a singlet pair clearly decays as $r^{-2}$ (dashed black line) at large $r$, which is consistent with the observations from DMRG. Furthermore, the probability distribution of bond energies at late times in the decimation (when only 50 free spins remain) approaches a power-law distribution agreeing with analytical predictions on the vacancy-free Heisenberg chain \cite{Fisher1994}, as shown in Fig. \ref{fig:numerical_rg}b. Our numerical SDRG analysis thus indicates that a disordered system with a finite density of localized holes still exhibits random-singlet-like behavior, corroborating the results from DMRG and the intuitive picture offered by the large-interaction limit of the Bethe Ansatz solution. 
\section{Discussion}

We have explored the ground state properties of Hubbard chains in the presence of quenched bond and site randomness, both at and away from half-filling. We find in all cases that disorder localizes charges and gives rise to random antiferromagnetic spin interactions, ultimately driving the system to a 
random-singlet-like phase. These results are consistent with the simple picture offered by the large-interaction limit of the 
Bethe Ansatz solution for the 
Hubbard chain, in which charges and spins are decoupled and respond independently to disorder. 


Our analysis here is specific to one dimension. In higher dimensions, one has to also consider the nontrivial effects of lattice geometry, particularly geometric frustration. Studies of (quasi) two-dimensional disordered spin systems suggest that geometry, alongside disorder, plays an important role in determining the spin state \cite{Laflorencie2005, Liu2018, Ren2020}; in this case, possibilities include short-range antiferromagnetic order, 
random-singlet, and spin glass order. 

Since all single particle states are localized in the random Hubbard chain,
the systems we have considered here offer valuable insight to the nature of the interacting insulator. For the same reason, these models do not allow us to make contact with the physics of local moments on the metallic side, or even near the metal-insulator transition. While most one-dimensional models suffer from the same affliction, higher-dimensional models might allow for study of metal-insulator transitions, but they prove significantly more difficult to solve without employing physically-motivated approximations\cite{Milovanic1989,Tusch1993, Pezzoli2010}.  Fortunately, one dimensional models with quasiperiodicity as a proxy for disorder, exhibit single-particle mobility edges\cite{DasSarma1990, DasSarma1988, Ganeshan2015}, and remain solvable in the presence of interactions\cite{Li2015}.  Thus, the extent to which their low energy behavior carries over to their disordered, higher dimensional counterparts, is an open question that can be investigated with a fair degree of rigor.  We shall report on this in future studies.

\section{Acknowledgments}
We thank V. Dobrosavljevic, S. Kivelson, P. Kumar, E. Miranda, C. Murthy, P. Nosov,  and J.-H. Son   for helpful discussions.  
SR and HCJ were supported by the Department of Energy, Office of Basic Energy Sciences, Division of Materials Sciences and Engineering, under Contract No. DE-AC02-76SF00515. JJY was supported by the National Science Foundation Graduate Research Fellowship under Grant No. DGE-1656518. RN acknowledges support from the Simons Foundation through a Simons Fellowship in Theoretical Physics, and from the Sloan Foundation through a Sloan Research Fellowship. Some of the computing for this project was performed on the Sherlock cluster. 

\section{Appendix}

\subsection{System with a random potential at half-filling}
\begin{figure}[h]
\begin{center}
\includegraphics[scale=1]{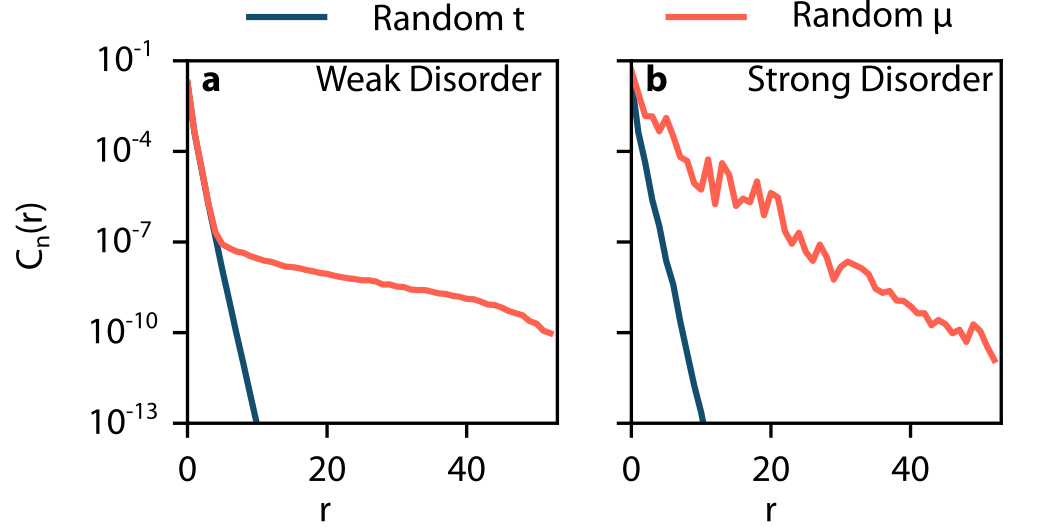}
\end{center}
\caption{Density-density correlation functions in half-filled Hubbard chains of length $L=72$ with (a) weak and (b) strong site and bond disorder. In both cases, density correlations in the chain with random hopping remain dominated by interactions and exhibit little change as the disorder is increased. In contrast, the chain with random potential undergoes a transition in behavior as the disorder increases. The density fluctuation correlations in the chain with weak random potential appear to have a short-range, interaction dominated region joined to a disorder-dominated region at large distances. At strong disorder, the same system becomes dominated by disorder.\label{fig:SUPhalffilling}}
\end{figure}

At half-filling, the clean Hubbard chain is a Mott insulator due to the presence of strong interactions, which force the ground state to have only singly-occupied sites. At half-filling, particle-hole symmetry of the Hubbard model allows for an interesting distinction between the effects of random hopping and random potential. The difference is demonstrated in the behaviors of the charge density-density correlation functions (Fig. \ref{fig:SUPhalffilling}a) at weak and strong disorders. While the weak random hopping model is still interaction-dominated and thus remains similar to the non-random half-filled Hubbard model, the weak random chemical potential model appears to promote charge density fluctuation correlations. For a range of $W_\mu$ sufficiently small, the density fluctuation correlations appear to have contrasting short- and long-distance behaviors. At sufficiently strong disorder, both the random potential and random hopping models yield exponentially decaying charge density-density correlations, although with different correlation lengths. 

This weak-to-intermediate disorder behavior of the half-filled chain may be understood by analyzing the negative-$U$ Hubbard model. The particle-hole transformation 
\begin{equation}
   c_{i,\uparrow}^\dagger \rightarrow (-1)^i d_{i,\uparrow} \quad , \quad c_{i,\downarrow} \rightarrow d_{i,\downarrow},
    \label{eq:ph_trans}
\end{equation}

maps the original Hubbard Hamiltonian in Eq.\ref{eq:hubbard_hamiltonian} to 
\begin{equation}
    \tilde{H} = -\sum_{i, \sigma}t_i \left(d_{i,\sigma}^\dagger d_{i+1,\sigma} +\text{h.c.} \right) - \sum_i \mu_i S_i^z  -U\sum_i \tilde{n}_{i\uparrow}\tilde{n}_{i\downarrow}, 
    \label{eq:ph_hubb}
\end{equation}
where $\tilde{n}_{i,\sigma} = d_{i,\sigma}^\dagger d_{i,\sigma}$. Comparing Eq. \ref{eq:ph_hubb} and Eq. \ref{eq:hubbard_hamiltonian}, we see that the spin and charge sectors effectively swap roles ($n_i \rightarrow S_i^z$) and that the sign of the interaction has flipped ($U\rightarrow -U$). The disorder in hopping remains bond disorder, but the disorder in chemical potential is transformed to a random magnetic field. Correspondingly, the charge density-density correlator (Eq. \ref{eq:densitydensity}) of the $U>0$ model is transformed into a spin-spin correlator $\expval{S_z(x)S_z(x+r)}$ in the $U<0$ model.  

In the large interaction limit $|U/t|\gg 1$, the ground state of the $U<0$ Hubbard model is superconducting. The $s$-wave superconducting state has a gap to spin excitations (interaction-dominated) and is stable to bond disorder. However, it is unstable to magnetic disorder. In the presence of weak magnetic disorder, the superconducting state retains the interaction-dominated behavior at short distances but shows evidence of pair-breaking at longer lengths, shown in Fig. \ref{fig:SUPhalffilling}a. For stronger magnetic disorder (Fig. \ref{fig:SUPhalffilling}b), the spins of the $U<0$ problem are completely locally polarized by the strong random fields, meaning the charges of the $U>0$ problem are localized by deep wells and high barriers in the random potential.

\subsection{Numerical SDRG Of the Hubbard Chain}
To corroborate the results from our DMRG calculations, we implement the real-space strong-disorder renormalization group (SDRG) procedure used  to characterize the ground state and infinite-randomness fixed point of the random antiferromagnetic Heisenberg chain\cite{Fisher1994}. We cannot apply this procedure directly, since we start with a Hubbard chain. Rather, we tackle the problem using perturbation theory in $t/U$ to find the effective low-energy (spin) Hamiltonian corresponding to the system, assuming the charges are localized. The initial Hamiltonian is Eq. \ref{eq:hubbard_hamiltonian}.

At half-filling the procedure is straightforward as the physical system is a Mott insulator, so the low-energy description is a Heisenberg antiferromagnet with no vacancies. The spin interaction between the $i$th and $i+1$th spins is found at second order in $t/U$ as $J_{eff}^{i} = 4t_{i}^2/U$ in the random-hopping chain. In the random-potential chain, we perturb the eigenstates of $H_0 = H_U + H_\mu$ with the correction $V = H_t$ to second order. Starting with $\ket{\psi_0} = (\ket{\uparrow, \downarrow} - \ket{\downarrow, \uparrow})/\sqrt{2}$, we see that acting with $V$ brings this to two intermediate states:
\begin{align*}
\ket{\uparrow\downarrow, 0} \quad &\quad E = U+\mu_i-\mu_j \\ 
\ket{0, \uparrow\downarrow} \quad &\quad E= U+\mu_j-\mu_i
\end{align*} 
Then, the total second-order energy correction to the singlet state is 
\begin{align*}
\Delta E^{(2)} = -\frac{2t^2}{U+(\mu_i-\mu_j)} - \frac{2t^2}{U-(\mu_i-\mu_j)}
\end{align*}

From this, we see that the effective spin interaction between two neighboring sites is
\begin{align*}
J_{eff}^{ij} &= \frac{2t^2}{U+(\mu_i-\mu_j)} + \frac{2t^2}{U-(\mu_i-\mu_j)} \\
&= \frac{4t^2}{U} \left[ \frac{1}{1-(\Delta \mu_{ij}/U)^2)}\right],
\end{align*}
where $\Delta\mu_{ij} = \mu_i - \mu_j$. This interaction remains antiferromagnetic so long as $W_\mu$ never exceeds $U$.

Away from half-filling, the evidence from DMRG suggests that the charges are still localized. We assume then that the holes sit at the local maxima of the chemical potential, and we can again recover a description the low-energy physics in terms of purely spin. In this case, spin interactions of neighboring spins with no vacancy separating them still take the form described above, but the spin interaction $J_{eff}^i$ between spins separated by a hole, at sites $i$ and $i+2$, is now found at fourth order in $t/U$ using Rayleigh-Schrodinger perturbation theory. Note that we must go to fourth order because the singlet and triplet energies are split only when the intermediate states involve double occupancy.  

In general, the fourth-order correction to the $n$-th energy in perturbation has the form: 
\begin{equation}
   \Delta E_n^{(4)} = \sum_{i\neq n}\sum_{j\neq n}\sum_{k\neq n} \frac{\bra{n} V \ket{i} \bra{i} V \ket{j} \bra{j} V \ket{k}\bra{k}V\ket{n} }{(E_n^{(0)} - E_k^{(0)})(E_n^{(0)} - E_j^{(0)})(E_n^{(0)} - E_i^{(0)})}.
\end{equation}

We consider a three-site system with a hole on site 2 (by construction, this implies $\mu_2>\mu_1, \mu_3$). Starting with the singlet state $\ket{S_{13}} = (\ket{\uparrow, 0, \downarrow} - \ket{\downarrow, 0, \uparrow})/\sqrt{2}$ which has energy $E=\mu_1 + \mu_3$, one can identify six possible contributions to the fourth-order energy correction, shown in Table \ref{tab:pt}. Let $\ket{S_{ij}}$ be the singlet state between spins on sites $i$ and $j$ and $V_{ij} = \bra{i}V\ket{j}$. Again, we consider the case of a random potential, so $H_0 = H_U + H_\mu$ and $V=H_t$.

\begin{table}[]
    \centering
   \begin{tabular}{c|c|c|c|c}
      Int. state 1 & $V_{01}$ & Int, state 2  & $V_{12}$ & Int. state 3 \\
     \hline  
     $\ket{S_{12}}$, $E=\mu_1+\mu_2 $& $t$ & $\ket{0, \uparrow\downarrow, 0} $, $E=2\mu_2 +U$& $\sqrt{2} t$ & $\ket{S_{12}}$\\
     $\ket{S_{12}}$, $E=\mu_1+\mu_2 $& $t$ & $\ket{0, \uparrow\downarrow, 0} $ , $E=2\mu_2 +U$ & $\sqrt{2}t$  & $\ket{S_{23}}$\\
     $\ket{S_{12}}$ , $E=\mu_1+\mu_2 $& $t$ &$\ket{\uparrow \downarrow, 0, 0}$ , $E=2\mu_1 +U$& $\sqrt{2}t$& $\ket{S_{12}}$\\
     $\ket{S_{23}}$ , $E=\mu_3+\mu_2 $& $t$& $\ket{0, \uparrow\downarrow, 0} $, $E=2\mu_2 +U$& $\sqrt{2}t$& $\ket{S_{23}}$\\
     $\ket{S_{23}}$ , $E=\mu_3+\mu_2 $& $t$& $\ket{0, \uparrow\downarrow, 0} $, $E=2\mu_2 +U$ & $\sqrt{2}t$& $\ket{S_{12}}$\\
     
     $\ket{S_{23}}$, $E=\mu_3+\mu_2 $& $t$& $\ket{ 0, 0, \uparrow \downarrow}$ , $E=2\mu_3 +U$& $\sqrt{2}t$& $\ket{S_{23}}$
\end{tabular}
    \caption{Contributions involving double occupancy of virtual states in the fourth-order energy correction of a singlet state across a hole. }
    \label{tab:pt}
\end{table}

Defining $\mu_{ij} = \mu_i -\mu_j$, we can write the effective spin interaction between spin 1 and 3 as: 

\begin{align}
\begin{split}
    J_{eff}^{1,3} &= \frac{2t^4}{U^3} \bigg[\frac{1}{(\mu_{12}/U)^2\cdot (1+\mu_{31}/U)}\\
    &+\frac{1}{(\mu_{32}/U)^2\cdot (1-\mu_{31}/U)} \\
    &+  \frac{1}{(\mu_{12}/U)^2\cdot (1-\mu_{12}/U-\mu_{32}/U)} \\
    & +  \frac{1}{(\mu_{32}/U)^2\cdot (1-\mu_{12}/U-\mu_{32}/U)}\\
    &+  \frac{1}{(\mu_{12}\mu_{32}/U^2)\cdot (1-\mu_{12}/U-\mu_{32}/U)}\\
    &+  \frac{1}{(\mu_{12}\mu_{32}/U^2)\cdot (1-\mu_{12}/U-\mu_{32}/U)}\bigg]
\end{split}
\label{eq:jeff_1hole}
\end{align}

Note that because $\mu_2 > \mu_1, \mu_3$ by construction, the last four terms in the effective interaction are guaranteed to be positive. So long as $W_\mu <U$, all terms in $J_{eff}$ are positive and therefore $J_{eff}$ remains antierromagnetic. The expression in Eq. \ref{eq:jeff_1hole} holds for the case of $\ell=1$ holes in a row. If there are $\ell\geq 2$ holes in a row, we approximate the effective spin interaction between the spins sandwiching the holes by the correct order of magnitude in $t/U$: $U(t/U)^{2\ell}$. Note that the interactions also remain antiferromagnetic for the same reason.

Given the similarity in the effects of the random hopping and random potential away from half-filling, we implement the SDRG procedure only for the case of the random potential, but we expect no qualitative difference if considering the system with random hoppings. We use a $L=600$ site chain, with $n=11/12$ electron filling.

The SDRG decimation procedure then proceeds as described for a random Heisenberg antiferromagnetic chain with no vacancies. At a given step, say the strongest bond $J_j$ connects the $j$-th spin to the $j+1$th spin (note that these may not reside on sites $j$ and $j+1$ away from half-filling). This bond is decimated, as spins $j$ and $j+1$ are locked into a singlet, giving rise to an effective bond of $\tilde{J} = J_{j-1}J_{j+1}/(2J_{j})$. In order to reconstruct the probability $P_s(r)$ of forming a singlet bond with spins separated by $r$ sites, we track the spatial indices of the singlets that are formed at each step and find the distribution of their separations $r$ over many realizations. 

To find the distribution of couplings near the end of the decimation procedure, we consider the couplings of each chain when there are $50$ remaining free spins. These couplings are normalized by the energy scale at each SDRG step (i.e., by the largest bond in the system at each step).  Fig. \ref{fig:numerical_rg}b in the main text shows the resulting histogram of normalized couplings across many realizations for each chain length. 

\subsection{Typical spin correlations}
We analyze the \textit{typical} (rather than average) spin correlations, $\overline{\ln| C_\sigma(r)|}$, which has a long-distance behavior $\overline{\ln |C_\sigma(r)|}\sim r^{0.5}$ in the random singlet phase \cite{Fisher1994}. We find typical spin correlations $\overline{\ln |C_\sigma(r)|}\sim r^{p}$, with $p$ between $0.42$ and $0.48$ for the region $L/4 < r <L/2$.

\begin{figure}[h]
\begin{center}
\includegraphics[scale=1]{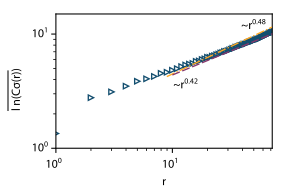} 
\end{center}
\caption{Typical spin correlations $\overline{\ln | C_\sigma(r)|}$ in a $L=144$ random potential chain, at $n=11/12$ electron filling, with $W_\mu = 3\bar{t}$.  \label{fig:avg_vs_sq}} 
\end{figure}

\bibliography{bibliography}

\end{document}